\documentclass{article}
\usepackage{amssymb,color,graphics}

\begin{document}
\newcommand{\beq}{\begin{equation}}
\newcommand{\eeq}{\end{equation}}

\title{Reply to comment of Bister and co-authors on the critique of the dissipative heat engine\footnote{$^1$Theoretical Physics Division, Petersburg Nuclear Physics Institute, Gatchina, St. Petersburg 188300, Russia;
$^2$Ecological Complexity and Modeling Laboratory, Department of Botany and Plant Sciences, University of California, Riverside, CA 92521-0124, USA;
$^3$Centro de Ciencia do Sistema Terrestre INPE, Sao Jose dos Campos SP 12227-010, Brazil}}

\author{A.M. Makarieva$^1$, V.G. Gorshkov$^1$, B.-L. Li$^2$, A.D. Nobre$^3$}

\label{firstpage}
\maketitle

\vspace{0.6cm}

\noindent
{\small {\it
I doubt not, but we have one common design; \newline
I mean, a sincere endeavour after knowledge...} \newline
Sir Isaac Newton}
\vspace{0.6cm}

\noindent
The dissipative heat engine (DHE) is based on a Carnot cycle with external heat $Q_{in}$
received at temperature $T_s$ and released at $T_o < T_s$. In contrast to the classical Carnot engine,
mechanical work $A_d$ in the DHE is not exported to external environment
but dissipates to heat within the engine.
Makarieva {\it et al.} (2010, hereafter MGLN) asserted that the laws of thermodynamics prohibit an increase of
$A_d$ beyond the Carnot limit: $A_d \le \varepsilon Q_{in}$, $\varepsilon \equiv (T_s - T_o)/T_s$.
Bister {\it et al.} (2010, hereafter BRPE) counterargued that such an increase is possible and
that hurricanes can be viewed as a natural DHE.
Here we show that the arguments of BRPE are not consistent with the energy conservation law and thus do not refute
MGLN's claims.

We first note that, unlike BRPE implied, MGLN did not confuse $A_d$ for work $W$ made on external bodies.
Indeed, that {\it the DHE does not produce any work on external environment} is obvious from the DHE definition
and leaves little space for confusion. MGLN consistently defined work $A_d$ as work dissipated within the engine.
Neither did MGLN characterize dissipation as a form of mechanical work. However, we do emphasize
that mechanical energy in the DHE dissipates at the same rate at which it is produced.
Thus, in BRPE's notations mechanical work $A_d$ produced by the DHE per unit time
is given by the unnumbered formula for dissipation rate $D$ on p. 3
(its relation to hurricane velocity $V$ is given by the unnumbered formula on p.~4 at $\beta = 0$):
$A_d = [(T_s - T_o)/T_o] Q_{in}$.
According to BRPE, while external work $W$ of any engine is bounded by the Carnot limit,
no limits exist to work $A_d$ and it can rise practically infinitely at very small $T_o$.

Any heat engine is put into operation by an external dynamic system that creates a departure from thermodynamic equilibrium
and determines the major parameters of the cycle. In our example (MGLN, Fig.~1) the piston is moved by a spring such that
the working gas can expand and receive heat isothermally.
In the hurricane the role of spring is played by the horizontal gradient of air pressure.
As MGLN illustrated, the first law of thermodynamics uniquely relates isothermal expansion of the gas to the amount of heat it receives.
If the expansion (set externally by the spring) does not change, the isothermal dissipation
of mechanical energy to heat within the DHE necessarily results in a decrease of the external heat input, while
the total amount of heat received by the gas remains constant. Consequently, work $A_d$
remains unchanged by dissipation instead of increasing as BRPE propose.
This can be seen in Fig.~1b of BRPE. Other things being equal, the electric heater put at the oceanic surface
warms the adjacent air and reduces the heat input from the ocean.
The cycle with electric heater (internal dissipation) shown in Fig.~1b of BRPE is then of equal intensity with the
ordinary Carnot cycle in Fig.~1a.

BRPE implicitly agree that to accomodate the {\it additional} heat input originating from dissipation
the gas must expand {\it further}. They refer to simulations of Bister \& Emanuel (1998, hereafter BE98) to claim that
"dissipative heating led to a further reduction in the central pressure, corresponding to enhanced isothermal expansion."
But to further reduce air pressure in the hurricane one has to perform additional work {\it on external environment}: indeed,
it is necessary to take the gas away from the hurricane and squeeze it to somewhere else.
The DHE {\it not performing any external work} cannot accomplish it.
The statement of BRPE that dissipative heating leads to a pressure reduction conflicts with the energy conservation law,
as the work needed to reduce the pressure originates then from nowhere.

Mass and momentum conservation and the first law of thermodynamics, from which BRPE claim BE98 derive their results,
all allow for solution $V = 0$ that corresponds to thermodynamic equilibrium and absence of air motion.
This is an inherent limitation of the thermodynamic approach.
Without identifying a physical mechanism that sustains a certain disequilibrium it is not possible
to describe hurricane dynamics. BE98 as well as Emanuel (1986, hereafter E86), whose work is the basis of BE98 approach, did so
by borrowing the essential parameters of the cycle from observations.
(For example, as the energy input to hurricane is supposed to be proportional to the difference of saturated and actual air enthalpies,
Eq. (1.1) of BRPE, and approaches zero at 100\% relative humidity, E86 and BE98 set relative humidity at around 80\%.
No theoretical explanation was given as to why, despite the proposed intense influx of
water vapour into the hurricane air, the air remains unsaturated.) Numerical simulations based on semi-empirical approaches
do not explain or predict the hurricane pressure reduction from the first physical principles and, as we illustrated above, may come
in conflict with the latter. There is an apparent need for new theoretical concepts in hurricane research; we hope that
the comment of BRPE and our reply will contribute to their development.

{\bf References}

Bister, M., Renn\'{o}, N. O., Pauluis. O. \& Emanuel, K. 2010 Comment on Makarieva et al. 'A critique of some
modern applications of the Carnot heat engine concept: the dissipative heat engine cannot exist'. Proc. R. Soc. A,
doi: 10.1098/rspa.2010.0087.

Bister, M. \& Emanuel, K. A. 1998 Dissipative heating and hurricane intensity. {\it Meteorol. Atmos. Phys.} {\bf 65,} 233-240.

Emanuel, K. A. 1986 An air-sea interaction theory for tropical cyclones. Part I. {\it J. Atmos. Sci.} {\bf 43,}
585-605.

Makarieva, A. M., Gorshkov, V. G., Li, B.-L. \& Nobre, A. D. 2010 A critique of some modern
applications of the Carnot heat engine concept: the dissipative heat engine cannot exist. {\it Proc.
R. Soc. A} {\bf 466,} 1893-1902.

\end{document}